\documentclass[aps,prl,twocolumn,showkeys,amsmath,amssymb,longbibliography,floatfix,superscriptaddress]{revtex4-1}
\usepackage[T1]{fontenc}
\usepackage{graphicx}
\usepackage{multirow}
\usepackage{dcolumn}
\usepackage{amsmath,amssymb,amsfonts,bm,dsfont,units}
\usepackage{hyperref}
\usepackage{tabularx}
\hypersetup{colorlinks=true,citecolor=blue,linkcolor=blue,urlcolor=blue,pdfstartview=FitH}
\usepackage{appendix} 
\usepackage{braket}
\usepackage{float,latexsym}
\usepackage{multirow}
\usepackage{diagbox}
\usepackage[normalem]{ulem} 
\usepackage[caption=false]{subfig} 
\usepackage{comment}
\usepackage{tikz}
\usepackage{xcolor}
\newcommand{\vect}[1]{\boldsymbol{#1}}
\usepackage{array}
\usepackage{graphicx} 
\usepackage{hhline,booktabs}
\usepackage{makecell}
\usepackage{wasysym} 
\usepackage{array}
\newcolumntype{Y}{>{\centering\arraybackslash}X}

\begin{document}
\title{The numerical case for identifying paired quantum Hall phases by their daughters
}

\author{Misha Yutushui}
\thanks{These authors contributed equally to this work.}
\affiliation{Department of Condensed Matter Physics, Weizmann Institute of Science, Rehovot 7610001, Israel}
 \author{Arjun Dey}
\thanks{These authors contributed equally to this work.}
 \affiliation{Laboratory for Theoretical and Computational Physics, PSI Center for Scientific Computing, Theory and Data, 5232 Villigen PSI, Switzerland}
\affiliation{Institute of Physics, \'{E}cole Polyetchnique F\'{e}d\'{e}rale de Laussane (EPFL), 1015 Laussane, Switzerland}

 \author{David F. Mross}
\affiliation{Department of Condensed Matter Physics, Weizmann Institute of Science, Rehovot 7610001, Israel}

\begin{abstract} 
Many candidate non-Abelian quantum Hall states are accompanied by nearby ‘daughter’ states, which are proposed to identify their topological order. Combining exact diagonalization and trial wave functions, we provide numerical evidence that daughter states reliably predict the parent topological phase. In the contexts of bilayer graphene and wide GaAs quantum wells, we show that the same interactions simultaneously stabilize Pfaffian, anti-Pfaffian, and their daughters, while suppressing the Jain states. The competition between Pfaffian and anti-Pfaffian, which is decided by particle-hole symmetry-breaking interactions, can likewise be deduced from their daughters. These findings strongly support the daughter-state-based identification of non-Abelian quantum Hall phases.
\end{abstract}

\date{\today}
\maketitle
\textit{Introduction.---}Non-Abelian particles are highly sought for their exotic properties~\cite{nayak_non-abelian_2008,Stern_non_abelian_2010}. Their many-body wave function encodes a quantum state together with its topological history~\cite{Jain_Incompressible_1989,Moore_nonabelions_1991,Wen_Non-Abelian_1991,Read_Beyond_1999}. As a result, controlled braiding of non-Abelian anyons could allow for intrinsically fault-tolerant manipulation of quantum information~\cite{Freedman_modular_functor_2000,Hormozi_Topological_quantum_2007}. Fractional quantum Hall states~\cite{Tsui_fqh_1982,Laughlin_fqh_1983,Haldane_fqh_1983} at specific filling factors are currently the leading platforms for realizing these particles. The simplest among them are Ising anyons, which are expected to occur in a half-filled Landau level of electrons~\cite{Morf_transition_1998,Park_possibility_1998,Rezayi_incompressible_2000,Scarola_Cooper_2000}. Half-filled plateaus have been observed in various platforms, including GaAs heterostructures~\cite{Willett_observation_1987,Eisenstein_Collapse_1988,Suen_Observation_1992,Wang_even_3_4_2022}, mono- and multi-layers of graphene~\cite{Ki_bilyaer_graphene_2014,Kim_bilayer_graphene_2015,Zibrov_Even_Denominator_2018,Kim_Even_Denominator_f_wave_2019,chen_tunable_2023,chanda_Even_TLG_2025}, and other materials~\cite{Falson_Zno_2015,Shi_even_wse2_2020}.

The experimental identification of non-Abelian quantum Hall states represents a significant challenge. It requires sensitivity to neutral modes that do not directly respond to conventional probes~\cite{Banerjee_observation_2018,Dutta_Isolated_2022,Paul_Thermal_2024,Melcer_Heat_2024,Bid_Observation_neutral_2010,Dutta_Distinguishing_2022}. While various experimental protocols have been proposed to determine topological order~\cite{Yutushui_Identifying_2022,Manna_Full_Classification_2022,Park_Fingerprints_2024,Yutushui_Identifying_2023,Yutushui_Universal_2025}, their implementation remains challenging or inconclusive~\cite{Simon_equilibration_2018,Feldman_comment_2018,Simon_reply_comment_2018,Feldman_equilibration_2019,Simon_equilibration_2020,Asasi_equilibration_2020}. Alternatively, multiple recent experiments have used Levin-Halperin daughter states~\cite{Levin_collective_2009} as indicators for the topological order at even denominator states~\cite{Singh_topological_2023,Huang_Valley_bilayer_graphene_2022,Kumar_Quarter_2024}. These daughter states arise when pairs of composite fermions form a bosonic integer quantum Hall state~\cite{Yutushui_daughters_2024,Zheltonozhskii_daughters_2024}. Under the assumption that the pairing is the same as in the parent even-denominator state, the daughter-state filling factors can uniquely determine the parent topological order. 

The validity of this assumption has not been tested experimentally or numerically. Moreover, daughter-state filling factors coincide with those of the ubiquitous Jain states at $\nu=\frac{n}{2n+1}$, and the topological order of the observed plateaus has not been established. They have been identified as daughter states by being anomalously strong and violating the monotonic decay of the Jain-state gap with $|n|$~\cite{halperin_theory_1993,Jain_composite_2007}. 

In this letter, we construct trial wave functions and use exact diagonalization techniques to test the relationship between even-denominator states, their daughters, and the Jain states. For two families of Hamiltonians, we find that the strengths of parents and daughters are well correlated, indicating their common origin. Moreover, they exhibit a very similar competition against states of unpaired composite fermions, i.e., the composite Fermi liquid (CFL) and the Jain states. Finally, we find that particle-hole symmetry-breaking interactions~\cite{Wojs_3body_2005,wang_particle_hole_2009,Wojs_landau_level_2010,Rezayi_breaking_2011,Pakrouski_phase_2015,Rezayi_Landau_2017}, which lift the degeneracy between Pfaffian and anti-Pfaffian, enhance or suppress the parent topological orders and their specific daughters simultaneously.

\begin{figure*}[t]
 \centering
 \includegraphics[width=0.99
 \linewidth]{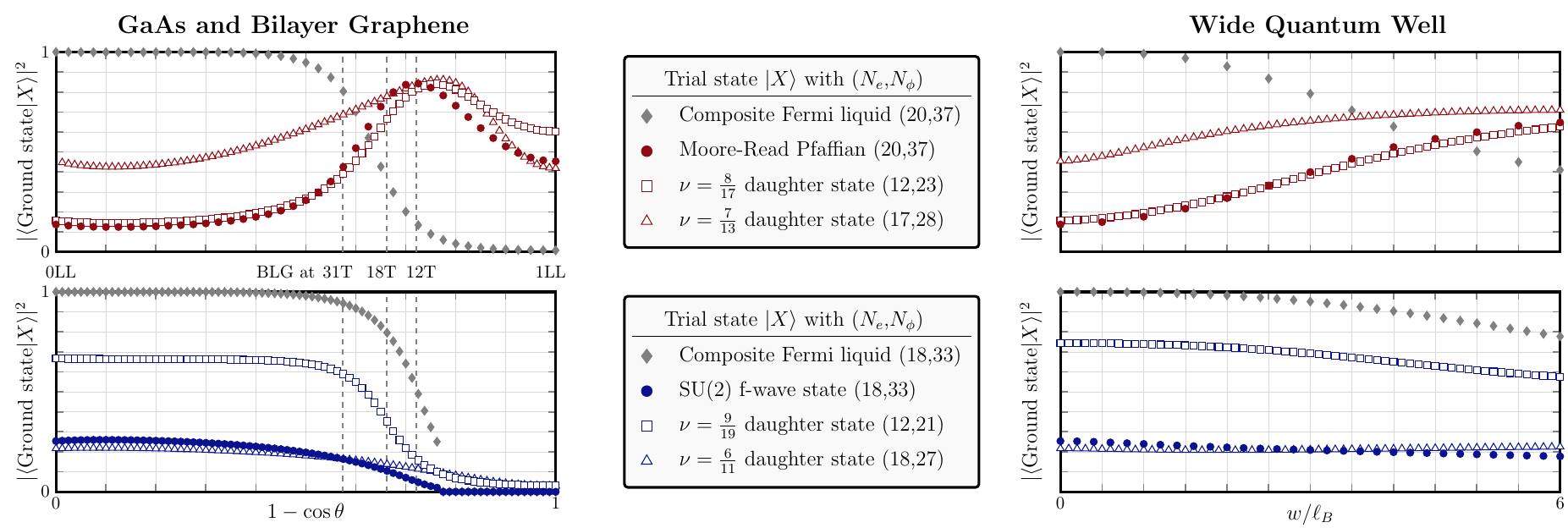}\\
 \caption{{\bf Paired states and their daughters in graphene and GaAs.} Left panel: The two lowest orbital Landau level wavefunctions of GaAs and bilayer graphene (BLG) can be parameterized as $\Psi=(\psi_0\cos \theta,\psi_1 \sin \theta ) $ for different choices of $\theta$. When the parameter $\theta \approx 0$, Coulomb interactions favor a composite Fermi liquid at $\nu=\frac{1}{2}$. For $\theta \approx \textstyle{\frac{\pi}{2}}$, we instead find that the Pfaffian or anti-Pfaffian states and their daughters are favored. By contrast, f-wave pairing and the corresponding daughters are not favorable for any $\theta$. Right panel: For interactions modelling a wide quantum well, we again find that the Pfaffian state and its daughters become simultaneously favorable with increasing width $w$, while the composite Fermi liquid is suppressed. States with f-wave pairing or its daughters are never favored. } 
 \label{fig.gaas}
\end{figure*}

\textit{Trial wave functions.---}As the primary diagnostic of quantum Hall states, we use the overlap between the ground state of a microscopic Hamiltonian and parameter-free model wave functions. For paired states and Jain states, trial wavefunctions based on the composite-fermion or parton paradigms are well established~\cite{Jain_Incompressible_1989,Moore_nonabelions_1991,Read_paired_2000,Jain_composite_2007,Balram_parton_2018,Yutushui_Large_scale_2020}. The daughter states at $\nu=\frac{6}{11}$ and $\frac{6}{13}$ have integer shifts ${\cal S} = 6$ and $-2$, respectively, and admit the parton representations~\cite{Balram_6_13_2018,Yutushui_daughters_2024}
\begin{align}\label{eq.wf_daughter_parton}
 \Phi_\frac{6}{11} = P_\text{LLL} \left[\phi_3\phi_2\phi_1\right],\qquad
 \Phi_\frac{6}{13} = P_\text{LLL} \left[\phi^*_3\phi^*_2\phi_1^3\right],
\end{align}
where $P_\text{LLL}$ is the lowest Landau projection and $\phi_n$ is the wavefunction of $n$ filled Landau levels.

The other pertinent daughter states do not permit wave functions of this type due to a fractional \textit{shift} ${\cal S}$~\cite{Wen_shift_1992}. States at $\nu=\frac{5}{11}$ and $\frac{7}{13}$ can be obtained by applying a particle-hole transformation to the wave functions in Eq.~\eqref{eq.wf_daughter_parton}. For $\nu = \frac{9}{17}$ and the $\nu=\frac{9}{19}$ daughter of the SU(2)$_2$, this option is not available, and we employ the anyon condensation approach~\cite{hansson2017quantum,yutushui_non_abelian_2025}. Specifically, for the $\nu=\frac{9}{17}$ daughter, we place \textit{pairs} of the quasiholes of a $\nu=1$ integer state into a bosonic $\nu=\frac{2}{9}$ Jain state $\Phi_{\frac{2}{9}} = \phi_1^5 \phi^*_2$, resulting in the wavefunction 
\begin{align}\label{eq.wf_clq}
 \Phi_{\frac{9}{17}}(z) = \phi_1(z)\int\limits_{u_a}\prod_{i}^{N_e}\prod^{N_{q}}_{a}(z_i-u_a)^2 \Phi^*_{\frac{2}{9}}(u_a).
\end{align} The integral is non-zero for $N_q=\frac{2}{9}(2N_e+3)$ quasihole pairs, correctly reproducing the quantum numbers of the $\nu=\frac{9}{17}$ daughter state. For the $\nu=\frac{9}{19}$ daughter, we replace the integer state by a $\nu=\frac{1}{3}$ Laughlin state $\phi_1 \rightarrow \phi_1^3$ and complex conjugate the integral. We further multiply the wavefunction by $\prod_{i<j}|z_i-z_j|^2$, which improves numerical convergence without affecting the topological phase. 

The resulting phases are Abelian topological orders described by a charge vector $q=(1,0,0)$ and $K$-matrices
\begin{align}
K_{\frac{9}{17}}=
 \begin{pmatrix} 1 & -2 & -2 \\
 -2 & - 4 & - 5\\
 -2 & - 5 & -4
 \end{pmatrix}, \quad
 K_{\frac{9}{19}}=
 \begin{pmatrix} 3 & -2 & -2 \\
 -2 & 4 & 5\\
 -2 & 5 & 4
 \end{pmatrix}.
\end{align}
Both matrices have two positive and one negative eigenvalues, implying a thermal Hall conductance of $\kappa_{xy}=1$ (in units of $\frac{\pi^2 k_B^2}{3h}T$), consistent with daughter states but not with Jain states at those fillings. Finally, using the spin vectors $s_{9/17}=\left(\frac{1}{2},1,2\right)$ and $s_{9/19}=\left(\frac{3}{2},1,2\right)$ to compute the shifts as ${\cal S}=\frac{2}{\nu}\vect{s} K^{-1} \vect{q}$, we obtain the expected values for the daughter states, i.e., $-\frac{1}{3}$ and $\frac{13}{3}$~\cite{Yutushui_daughters_2024}.

We obtain second-quantized representations of the wave functions in Eq.~\eqref{eq.wf_daughter_parton} and \eqref{eq.wf_clq} by computing their overlaps with a full set of basis states in the corresponding Hilbert space via Monte-Carlo sampling; see Appendix~\ref{sec.anyonmc} for details. Similarly, we construct the $\nu=\frac{2}{5},\frac{3}{7}$ Jain states and the SU(2)$_2$ parton states. For the $\nu=\frac{1}{3}$ state and the Pfaffian, we use Jack polynomials~\cite{Bernevig_Anatomy_2009}.

\textit{Daughter states in GaAs and graphene.---}We consider Coulomb interactions projected into one partially filled Landau level of single-electron orbitals of the form $\Psi(\theta)=(\psi_0\cos \theta,\psi_1 \sin \theta )$. Here, $\psi_n$ are the wave functions of non-relativistic free electrons in the $n$th Landau level. In particular, the cases $\theta=0$ and $\theta=\frac{\pi}{2}$ describe the lowest and first excited Landau levels of GaAs. Previous numerical studies \cite{Morf_transition_1998,Park_possibility_1998,Rezayi_incompressible_2000,Scarola_Cooper_2000,Wojs_3body_2005,wang_particle_hole_2009,Wojs_landau_level_2010,Rezayi_breaking_2011,Pakrouski_phase_2015,Rezayi_Landau_2017}
found that Coulomb interactions in the half-filled lowest Landau level favor a composite Fermi liquid and Pfaffian or anti-Pfaffian pairing in the first excited Landau level.

In bilayer graphene, the eight-fold degenerate zeroth Landau level contains four $\theta=0$ levels and four with an intermediate $\theta$ that depends weakly on the magnetic field and displacement field. The values range between $\theta_{12T} \approx 0.41 \pi $ at $B=12T$ and $\theta_{31T} \approx 0.36\pi $ at $B=31T$ under typical experimental conditions~\cite{hunt_direct_2017,Haug_Interaction_2025}. 

We use the DiagHam library~\cite{DiagHam} to exactly diagonalize $H(\theta)$ in a spherical geometry. For $N_e=20$ particles and $N_\phi=37$ magnetic flux quanta, a compressible composite Fermi liquid and an incompressible state with Pfaffian pairing can both be realized. Fig.~\ref{fig.gaas} shows that the ground state transitions from a composite Fermi liquid at $\theta \lesssim 0.37\pi $ to a paired state, with the largest overlap for $\theta \approx \theta_{12T}$. 
\begin{figure}[t]
 \centering
 \includegraphics[width=0.99
 \linewidth]{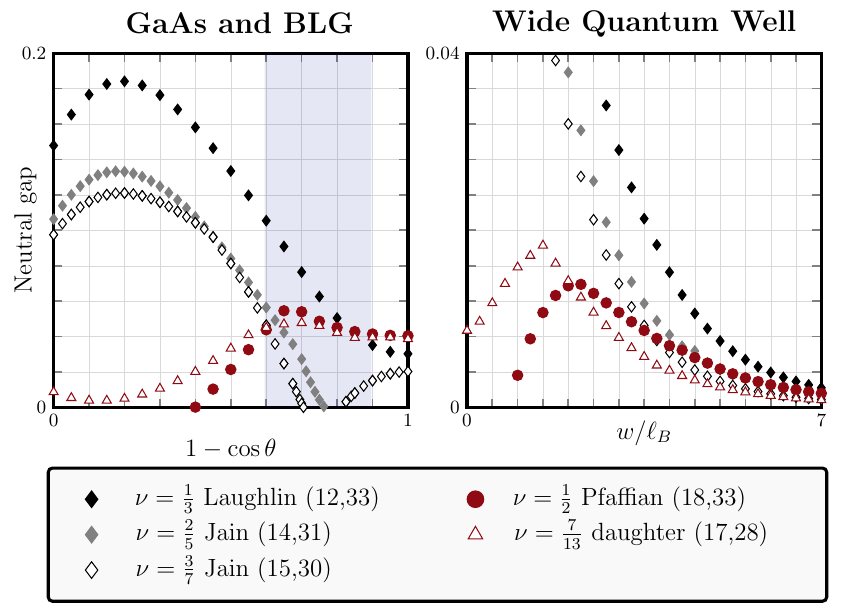}
 \caption{{\bf Competition between Jain states and daughter states.} We numerically compute the lowest-energy eigenstates for systems whose $(N_e,N_\phi)$ match the indicated states. When the ground state has angular momentum zero, we compute the energy difference to the first excited states. As anticipated, we find that this neutral gap is suppressed more quickly with $\theta$ or $w$ for higher-order Jain states, such that the daughter state becomes competitive. The shaded region in the left plot indicates where Ref.~\cite{Balram2025daughter} found an $L=0$ ground state in a $(N_e,N_\phi)=(20,40)$ system, which corresponds to the $\frac{8}{17}$ daughter and is consistent with our findings. }
 \label{fig.gaps}
\end{figure}
The $\nu=\frac{7}{13}$ daughter of the Moore-Read Pfaffian occurs for $N_e=17$, $N_\phi=28$. As its trial state, we use the particle-hole conjugate of the $\nu=\frac{6}{13}$ daughter state defined in Eq.~\eqref{eq.wf_daughter_parton}. We find that its overlap with the $H(\theta)$ 
ground state also exhibits a pronounced maximum near $\theta_{12T}$. The overlap at $\theta \approx 0$ is significant, which we attribute to the absence of a strong competitor, i.e., the daughter state at this particle number is not aliased with (does not reside in the same Hilbert space as) any Jain state; see Appendix~\ref{sec.hilbert}. 

The $\nu=\frac{8}{17}$ daughter of the Pfaffian is the particle-hole conjugate of $\Psi_\frac{9}{17}$ in Eq.~\eqref{eq.wf_clq}, and both $\Psi_\frac{8}{17},\Psi_\frac{9}{17}$ can occur in the $N_e=12$, $N_\phi=23$ system, i.e., the Pfaffian daughter is aliased with its own particle-hole conjugate. The eigenstates of any two-body Hamiltonian in this system have a well-defined transformation under particle-hole transformation, which is even for the $H(\theta)$ ground state. Consequently, we compare it to the equal superposition of our $\nu=\frac{8}{17}$ and $\nu=\frac{9}{17}$ trial states. The resulting overlap in Fig.~\ref{fig.gaas} closely tracks the ground state overlap of the Pfaffian.

To test whether this behavior is indeed related to the specific pairing channel, we repeat this analysis for the SU(2)$_2$ state and its two daughters. We find that all of these states are strongly suppressed as $\theta$ increases. These results support the hypothesis that the experimental observations of $\nu=\frac{7}{13}$ and $\nu=\frac{8}{17}$ daughter states (or their particle-hole conjugates) point to the Pfaffian (or anti-Pfaffian) topological order at half-filling.

\textit{Daughter states in wide quantum wells.---}In wide GaAs quantum wells, the transverse extent $w$ of the well represents a continuous tuning parameter. A non-zero width reduces the energetic penalty for electrons at small in-plane separation and can promote pairing at even-denominator fillings of the lowest Landau level. In practice, an in-plane magnetic field or tuning the electron density with respect to the transverse sub-bands permits in-situ control over $w$. 

To model the wide quantum well, we replace the Coulomb interactions by $V_w(r) = \frac{1}{\sqrt{r^2 + w^2}}$~\cite{Zhang_Excitation_1986}. As before, we numerically obtain the ground states on a sphere for different choices of $(N_e,N_\phi)$ and compute the overlaps with the same trial states. The results are shown in the right panel of Fig.~\ref{fig.gaas}. With increasing $w$, the composite Fermi liquid is suppressed, while the Pfaffian order and its two daughters are favored; consistent with the experimental observations of Ref.~\cite{Singh_topological_2023}. The f-wave paired state and its two daughters are not favorable at any $w$.

\textit{Competition between Jain and daughter states.---}The results presented above support the assertion that the presence of specific daughter states points to the parent topological order. We thus turn to the question of whether the observed plateaus in bilayer graphene and wide quantum wells are indeed daughter states or merely Jain states. To address this issue numerically, we assess the strengths of different topological phases as $\theta$ or $w$ are varied. Specifically, for any $L=0$ ground state, we compute the gap $\Delta$ to the first excited state in the same Hilbert space, focusing on systems where Jain states, paired states, and daughter states do not directly compete. Unfortunately, an extrapolation of the gaps to the thermodynamic limit is not possible due to the small number of accessible system sizes. To increase the system size at filling $\nu=\frac{q}{p}$, one needs to increment $(N_e,N_\phi) \rightarrow (N_e+q,N_\phi+p)$---a vast increase in Hilbert space size at daughter state fillings (cf.~Appendix~\ref{sec.hilbert} for Hilbert space dimensions). Moreover, all daughter states apart from the one at $\nu=\frac{7}{13}$ (and its particle-hole conjugate) are aliased with Jain states at available system sizes. We thus restrict our study of gaps to the $\nu=\frac{7}{13}$ daughter, the first three Jain states, and the Pfaffian.

Fig.~\ref{fig.gaps} shows the numerically obtained gap whenever the ground has zero angular momentum. As expected, we find that all Jain states are suppressed with increasing $\theta$ or $w$, and higher-order Jain states are affected more strongly \footnote{The re-emerging $L=0$ states for $(15,30)$ near $\theta=\frac{\pi}{2}$ have negligible overlap ($0.1\%$) with those at smaller $\theta$. They likely describe a distinct phase from the Jain $3/7$, but the filling and shift do not match any other topological order known to us.}. In contrast, the $\nu=\frac{7}{13}$ state and Pfaffian are strengthened at intermediate values and can become favored even over low-order Jain states. These findings support the claim that a well-developed quantum Hall state at $\frac{6}{13}$ (or $\frac{7}{13}$) with much weaker or absent states at lower Jain fillings, such as $\frac{5}{11},\frac{4}{9}$ (or $\frac{6}{11},\frac{5}{9}$), is likely to be a daughter state.

\textit{Landau level mixing.---}All results presented so far were obtained for models with pure two-body interactions in a single Landau level. For such interactions, any quantum Hall state at filling factor $\nu$ is degenerate with its hole-conjugate at filling $1-\nu$. At half-filling, different pairing channels such as the Pfaffian and anti-Pfaffian orders are of identical strength and must likewise be degenerate.

In experiments, particle-hole symmetry is broken due to Landau-level mixing. When the mixing is weak, it can be included perturbatively through a modified interaction between electrons in a single Landau level. At the leading order, Landau-level mixing renormalizes the two-body interactions, which preserve particle-hole symmetry, and introduces three-body interactions, which break it. Numerous studies used such interactions to assess the competition between Pfaffian and anti-Pfaffian in GaAs at $\nu=\frac{5}{2}$~\cite{Ma_Fractional_5_2_2022}. Instead of revisiting this debate, we focus on the most pertinent question in the present context: Is the realization of Pfaffian daughters instead of anti-Pfaffian daughters indicative of the topological order at half filling?

As a first test, we use the well-known parent Hamiltonian of the Pfaffian, $H_3= \sum_{i<j<k}{\cal P}_{ijk}(3)$, where ${\cal P}_{ijk}(\ell)$ projects the electron triplet $i,j,k$ onto a state with relative angular momentum $\ell$. These interactions have particle-hole symmetric contributions, which are expressible by two-body pseudopotentials \cite{Peterson_spontaneous_2008,Sreejith2017,kusmierz2019,Hutzel2019} and anti-symmetric components. Using the $H_3$ at the shift of the $\nu=\frac{7}{13}$ daughter state, we find $\approx 84\%$ squared overlap of the $N_\phi=27$ ground state with the Pfaffian daughter trial wavefunction. By contrast, the $\nu=\frac{6}{13}$ ground state exhibits an overlap below $1\%$ with the anti-Pfaffian daughter. 

Encouraged by this finding, we systematically perturb the particle-hole symmetric Hamiltonian $H(\theta)$ studied above by a particle-hole \textit{antisymmetric} interaction 
\begin{align}
 \delta H_\ell\equiv \sum_{i<j<k}[{\cal P}_{ijk}({\ell})-\bar{\cal P}_{ijk}({\ell})].
\end{align}
For $\ell=3$ or 5, the resulting $\delta H_\ell$ includes not only the three-body potential, but also the first and third (and fifth) two-body pseudopotentials, whose strengths depend on the number of magnetic flux quanta piercing the sphere; see Table~\ref{tab.pseudos}. We then determine ground-state overlaps and the many-body gap of $H_\text{total}=H(\theta_{18T}) + \gamma \delta H$, where $\gamma=0$ realizes the Pfaffian or its daughters according to Fig.~\ref{fig.gaas}. 

\begin{figure}[t]
 \centering
 \includegraphics[width=0.99
 \linewidth]{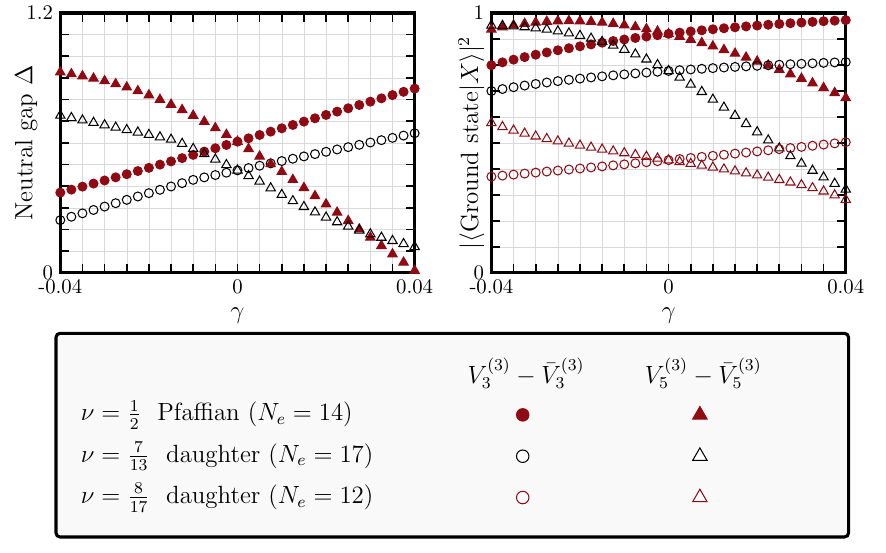}\\
 \caption{
 {\bf Particle-hole symmetry breaking.} We study Landau-level mixing by perturbing the $\theta=\theta_{18T}$ Coulomb Hamiltonian with the particle-hole antisymmetric part of the third and fifth three-body pseudopotentials. The neutral gaps $\Delta$ (left) and the ground state overlaps (right) of the Pfaffian and its $\nu=\frac{6}{13}$ daughter closely track one another for both perturbations. For $\nu=\frac{8}{17}$, we only show the overlap but no gap, since this state is aliased with its own particle-hole conjugate in the only available system, for which $N_e$=12.}
 \label{fig.llm}
\end{figure}
\begin{table}[t]

	\centering
	\renewcommand{\arraystretch}{1.2} 
	\centering
	\caption{{\bf Pseudopotentials} Antisymmetrizing a pure three-body Hamiltonian with respect to particle-hole leads to two-body pseudopotentials whose strength depends (weakly) on the number $N_\phi$ of magnetic flux quanta piercing the sphere.}
	\begin{tabular}{r|c c|c c c }
 \hline\hline 
		& \multicolumn{2}{c|}{ ${\cal P}_{ijk}({3})-\bar{\cal P}_{ijk}({3})$} & \multicolumn{3}{c}{${\cal P}_{ijk}({3})-\bar{\cal P}_{ijk}({3})$}\\
 		$N_\phi$ & $V^{(2)}_1$& $V^{(2)}_3$& $V^{(2)}_1$& $V^{(2)}_3$& $V^{(2)}_5$\\\hline
 23 &$-1.61240$ &$-0.57175$ &$-1.10720$&$ -0.30608 $& $-0.74666$\\
 25 &$-1.61854$ &$-0.57092$ &$-1.12006 $&$ -0.30389 $& $-0.74248$\\
 28 & $-1.62607$ & $-0.56992 $&$-1.13586$&$-0.30125$ &$ -0.73753 $ \\
		\hline\hline
	\end{tabular}
	\label{tab.pseudos}
\end{table}

When we perturb with $\delta H_3$, the many-body gap at the Pfaffian shift $(N_e,N_\phi)=(14,25)$ increases as expected for positive $\gamma$, as shown in Fig.~\ref{fig.llm} (left). Simultaneously, the overlap with the Pfaffian wave function increases; see Fig.~\ref{fig.llm} (right). For the system size $(N_e, N_\phi)=(17,28)$, appropriate for the $\frac{7}{13}$ daughter state, we observe a very similar evolution of the gap and overlap, i.e., an increase for positive $\gamma$. We do not show the gap of the second Pfaffian daughter at $\nu=\frac{8}{17}$, since it is aliased with its own particle-hole conjugate at the only accessible system size. Still, the overlap with the trial state follows the same trend as for the other two states. When perturbing with $\delta H_5$, we find that a positive coefficient suppresses gaps and overlaps for the Pfaffian and its two daughters. 

The gaps and overlaps for anti-Pfaffian and its two daughters behave exactly oppositely; they correspond to $\gamma \rightarrow -\gamma$ and are not shown separately. We interpret these results as evidence that the observation of daughter states can indeed distinguish between parent Pfaffian and anti-Pfaffian orders, independently of the precise particle-hole symmetry-breaking perturbation.

\textit{Conclusions.---}We have provided numerical evidence that the quantum Hall plateaus flanking half-filled states in bilayer graphene and wide GaAs quantum wells are Levin-Halperin daughter states. More generally, our study supports the case that the topological orders of paired quantum Hall states can be inferred from their daughters, including the distinction between Pfaffian and anti-Pfaffian. We find no indication of different pairing channels at half-filling in the zeroth Landau level of bilayer graphene and GaAs. Still, quantum Hall states with other pairing channels could also be diagnosed by their daughters; see Appendix~\ref{fwave} for a specific example, where two-body interactions stabilizing the SU(2)$_2$ topological order also favor its daughters.

{\it Data Availability}---The daughter states and parton SU(2)$_2$ trial wave functions generated by Monte-Carlo sampling for this study are available at \cite{Zenodo}, stored in DiagHam-compatible binary format.

\begin{acknowledgments}
\textit{Acknowledgments.}---It is a pleasure to acknowledge illuminating discussions with Ajit Balram, Maria Hermanns, and Yuval Ronen. 
This work was supported by the Israel Science Foundation (ISF) under grant 2572/21 and by the Minerva Foundation with funding from the Federal German Ministry for Education and Research. AD thanks the High Performance Computing and Emerging Technologies group
at PSI, which maintains the Merlin7 cluster, where part of the numerical work was done. 
\end{acknowledgments}

\appendix

\section{Monte-Carlo for anyon condensation states}
\label{sec.anyonmc}
To obtain a Fock-space representation of a real-space wave function such as Eq.~\eqref{eq.wf_clq}, we evaluate a large number of multi-dimensional integrals of the form
\begin{equation}
\begin{split}
 \langle \Phi_I | \Psi_\text{trial} \rangle &= \int_{z} \Phi_I(z^*) \Psi_\text{trial}(z)~\\
 &= \int_{z} \Phi_I(z^*) \Psi(z)\int_u (u-z)^k \Psi_\text{QP}(u^*)~.\label{eqn.overlaps}
\end{split}
\end{equation}
with integer $k$.
Here, the $N$ component vector $I=(i_1,i_2,\ldots,i_N)$ defines a basis state of the Fock space; in the spherical geometry
\begin{align}
 \Phi_I = \det \begin{pmatrix}
 Y^Q_{Q,m_{i_1}}(\vect r_1) & Y^Q_{Q,m_{i_1}}(\vect r_2) & \ldots\\
 Y^Q_{Q,m_{i_2}}(\vect r_1) & Y^Q_{Q,m_{i_2}}(\vect r_2) & \ldots\\
 \vdots& \vdots & \ddots~
 \end{pmatrix}
 ,
\end{align}
where $Y^Q_{l,m}$ are monopole Harmonics. The number of basis states grows exponentially with the system size, and evaluating all these Slater determinants is the numerical bottleneck. However, we note that changing the $u$ coordinate in Eq.~\eqref{eqn.overlaps} affects $\Psi_\text{trial}$ but not $\Phi_I$, i.e., all the determinants can be reused for sampling the overlap. To reduce the computational cost, we thus make $10^3$-$10^4$ $u$-updates in between $z$-updates.

To refine the wave functions obtained via Monte-Carlo sampling, we then project them onto the ${\vect L}^2 =0$ subspace as done in Ref.~\cite{Mishmash_numerical_2018}. By comparing the final wave functions resulting from independent runs with comparable numerical effort, we estimate that the squared overlaps of the numerically obtained wave functions with the exact ones exceed $99\%$.

\section{Haldane pseudopotentials for Bilayer graphene}
The zeroth Landau level of bilayer graphene contains four $N=0$ and four $N=1$ orbitals that are describable by two-component wave functions $\Psi = (\psi_0 \cos\theta, \psi_1\sin\theta)$ with zero and non-zero $\theta$, respectively. In spherical geometry, the $n$th Landau level has $2(Q+n)+1$ states, where $Q$ is the monopole strength. To have equal degeneracies for both $n=0$ and $n=1$ components, we thus use the monopole strengths $Q_n=Q-n$.

The orbitals in the zeroth Landau level of bilayer graphene are thus
\begin{align}\label{eq.BLG_Y}
 {\cal Y}^{Q}_{m} =(
 C_0 Y^{Q}_{Q,m} ,
C_1 Y^{Q-1}_{Q,m} ),
\end{align}
where we introduced the vector $(C_0,C_1) = (\cos \theta,\sin\theta)$. The matrix elements of a real-space density-density potential $V(r)$ are then computed as
\begin{align}
 {\cal V}_{m_1,m_2}^{m_3,m_4}(\theta) &= \int_{\vect r_1}\int_{\vect r_2} \; V(|\vect r_1-\vect r_2|)\nonumber\\
 & \times \left({\cal Y}^{Q\dag}_{m_1}(\vect r_1) \cdot {\cal Y}^Q_{m_3}(\vect r_1)\right)
 \left({\cal Y}^{Q\dag}_{m_2}(\vect r_2)\cdot {\cal Y}^Q_{m_4}(\vect r_2)\right)\nonumber\\
 & = \sum_{i =0,1}\sum_{j =0,1} [C_iC_j]^2 V_{m_1,m_2}^{m_3,m_4}(Q_i,Q_j).\quad \label{eqn.curlyV.second}
\end{align}
In the last row, we used the non-relativistic matrix elements for electrons in two different $i$th and $j$th Landau levels with monopole strengths $Q_i$ and $Q_j$, respectively, i.e.,
\begin{align} 
 V_{m_1,m_2}^{m_3,m_4}(Q;Q_i,Q_j) =
 \int_{\vect r_1}\int_{\vect r_2} \; V(|\vect r_1-\vect r_2|)\\
 \times Y^{Q_i*}_{Q,m_1}(\vect r_1)Y^{Q_i}_{Q,m_3}(\vect r_1)\;
 Y^{Q_j*}_{Q,m_2}(\vect r_2)Y^{Q_j}_{Q,m_4}(\vect r_2)\nonumber
\end{align}
Here, the number of states is $Q$, the same for both the $i$th and $j$th Landau levels. Eq.~\eqref{eqn.curlyV.second} involves these elements for $i,j=0,1$ Landau level indices.

For a rotationally symmetric potential, only matrix elements with $\Delta m=m_3-m_1=m_4-m_2$ are non-zero. Expanding the real-space potentials in terms of Legendre polynomials $V(\theta) = \sum_{k}V_k P_k(\cos(\theta))$, using the addition theorem of spherical harmonics and the integral of three monopole harmonics~\cite{Jain_composite_2007,wooten_haldane_2013,Yutushui_phase_2025,Balram_BLG_2022}, the matrix elements are readily found to be 
\begin{align}
V_{m_1,m_2}^{m_3,m_4}(Q;Q_i,Q_j) &=
 \sum_{k=0}^{2Q}V_k(2Q+1)^2 (-1)^{2Q-m_3-m_2}\\
 &\times \begin{pmatrix}Q&k&Q\\m_1& \Delta m & -m_3 \end{pmatrix}
 \begin{pmatrix}Q&k&Q\\-Q_i& 0& Q_i \end{pmatrix}\nonumber\\
 &\times \begin{pmatrix}Q&k&Q\\m_2& \Delta m & -m_4 \end{pmatrix}
 \begin{pmatrix}Q&k&Q\\-Q_j& 0& Q_j \end{pmatrix}.\nonumber
\end{align}
The Haldane pseudopotentials corresponding to $V_{m_1,m_2}^{m_3,m_4}(Q;Q_i,Q_j)$ are given by
 \begin{align} \label{eq.Hq1q2}
 &H^Q_{L}(Q_i,Q_j) = \sum_{m_i=-Q}^{Q}
 V_{m_1,m_2}^{m_3,m_4}(Q;Q_i,Q_j) \\
 &\times
 \sum_{M'}\langle 2Q-L, M'|Q,m_3;Q,m_4\rangle
 \langle Q,m_1;Q,m_2|2Q-L, M'\rangle\nonumber
 \\&=(2Q+1)^2 (-1)^{2Q-Q_i-Q_j-L}\label{eq.Haldane}\\
 &\times\sum_{k=0}^{2Q}
 V_k
 \begin{pmatrix}Q&k&Q\\-Q_i& 0& Q_i \end{pmatrix} 
 \begin{pmatrix}Q&k&Q\\-Q_j& 0& Q_j \end{pmatrix}
 \left\{\begin{matrix}2Q-L&Q&Q\\k& Q& Q \end{matrix}\right\}. \nonumber
 \end{align}

Finally, we obtain the bilayer graphene pseudopotentials as
\begin{align}
 {\cal H}^\theta_L = \sum_{i =0,1}\sum_{j =0,1} [C_iC_j]^2 H^{Q}_{L}(Q_i,Q_j).
\end{align}

\section{Daughter-state Hilbert spaces and aliasing}\label{sec.hilbert}
On a finite sphere, the particle number and flux required to realize a particular topological order in the ground state are determined by the filling factor and shift according to 
\begin{align}
 N_\phi = \nu^{-1} N_e - \cal {S}.
\end{align}
Table~\ref{tab.aliasing} summarizes the Hilbert spaces and their dimension for the four daughter states. Black rows indicate the systems studied in this work, while gray indicates the next system where the same orders could occur. The final column lists other states that can also be realized in the same Hilbert space.
\begin{table}[t]

	\centering
	\renewcommand{\arraystretch}{1.4} 
	\centering
	\caption{{\bf Finite-size Hilbert spaces of daughter states.} The filling factor $\nu$ and shift ${\cal S}$ determine the required number of flux quanta $N_\phi$ piercing the surface sphere for $N_e$ electrons. Here we list the dimensions of these Hilbert spaces, and indicate when the Hilbert spaces coincide with those of other daughter states, Jain states, or paired quantum Hall states. Systems listed in gray are included for completeness, but were not part of this study.}
	\begin{tabular}{c c | c c c c c}
 \hline\hline 
		$\nu$& ${\cal S}$ & $N_e$ & $N_\phi$& $D_{L_z=0}$ 
 & $D_{\vect{L}=0}$&Aliased states
 \\\hline 
 \multirow{ 3}{*}{$\frac{8}{17}$} &\multirow{ 3}{*}{$\frac{5}{2}$} & \multirow{ 2}{*}{$12$} & \multirow{ 2}{*}{$23$} & \multirow{ 2}{*}{$61108$} & \multirow{ 2}{*}{$127$} & $\nu=\frac{3}{7}$ Jain, PH-Pfaffian, \\
 &&&&&&$\nu=\frac{9}{17}$ daughter \\
 & &\textcolor{gray} {$20$} & \textcolor{gray} {$40$ }& \textcolor{gray} {$2.8*10^{9}$} & \textcolor{gray} {903256 }&\textcolor{gray} {$\nu=\frac{10}{19}$ daughter }\\
 \hline 
 \multirow{ 3}{*}{$\frac{7}{13}$} &\multirow{ 3}{*}{$\frac{25}{7}$} & \multirow{ 1}{*}{$17$} & \multirow{ 1}{*}{$28$} & \multirow{ 1}{*}{$901723$} & \multirow{ 1}{*}{$902$} & \\
 & & \multirow{ 2}{*}{\textcolor{gray} {$24$}} & \multirow{ 2}{*}{\textcolor{gray} {$41$}} & \multirow{ 2}{*}{\textcolor{gray} {$3.5*10^{9}$}} & \multirow{ 2}{*}{\textcolor{gray} {$1099354$}} & \textcolor{gray} {$\frac{3}{5}$ Jain},\\ 
 & & & & & & \textcolor{gray} {$\nu=\frac{1}{2}$ with $\ell=5$ pairing}\\\hline
 \multirow{ 3}{*}{$\frac{9}{19}$} &\multirow{ 3}{*}{$\frac{13}{3}$} & \multirow{ 2}{*}{$12$} & \multirow{ 2}{*}{$21$} & \multirow{ 2}{*}{$16660$} & \multirow{ 2}{*}{$58$} & $\nu=\frac{3}{5}$ Jain, $\frac{4}{9}$ Jain,\\
 &&&&&&Pfaffian \\
 & & \textcolor{gray} {$21$ }& \textcolor{gray} {$40$ }& \textcolor{gray} {$2.8*10^{9}$ }&\textcolor{gray} { $903256$ }&\textcolor{gray} { $\nu=\frac{9}{17}$ daughter }\\
 \hline 
 \multirow{ 3}{*}{$\frac{6}{11}$} &\multirow{ 3}{*}{$6$} & \multirow{ 2}{*}{$18$} & \multirow{ 2}{*}{$27$} & \multirow{ 2}{*}{$246448$} & \multirow{ 2}{*}{$319$} & $\nu=\frac{2}{3}$ Jain,\\
 &&&&&&$\nu=\frac{1}{2}$ with $\ell=7$ pairing \\
 & & \textcolor{gray} {$24$ }& \textcolor{gray} {$38$} & \textcolor{gray} {$2.9*10^8$} & \textcolor{gray} {114049}& \\\hline\hline
	\end{tabular}
	\label{tab.aliasing}
\end{table}
\section{F-wave paired states and their daughters}\label{fwave}At present, there is no strong candidate for an experimentally realized SU(2)$_2$ topological order. Numerical works have suggested such a state in monolayer graphene at~\cite{Kim_Even_Denominator_f_wave_2019,Sharma_unconventional_2022}, but found it to be rather fragile. Its stabilization requires a strong fifth Haldane pseudopotential, $V_5$ \cite{Yutushui_phase_2025}, which is not realized in low Landau levels where most half-filled states occur. Still, we test the generality of the daughter-state-based identification by studying a Coulomb Hamiltonian perturbed by $V_5$.

\begin{figure}[h]
 \centering
 \includegraphics[width=0.99
 \linewidth]{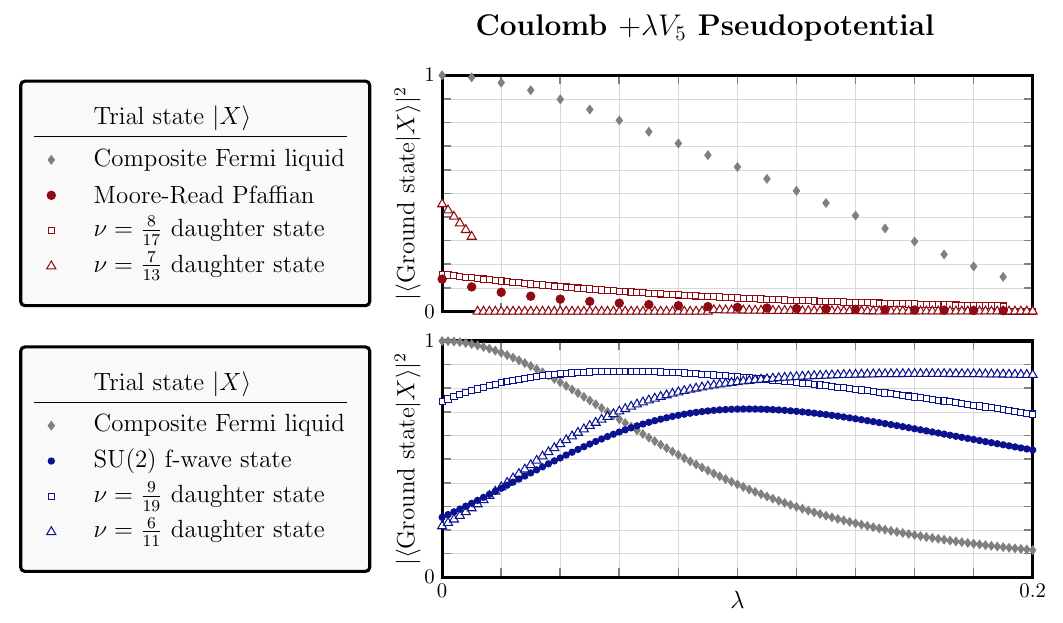}\\
 \caption{{\bf SU(2)$_2$ and its daughters for strong $V_5$ pseudopotential.} Perturbing Coulomb interactions in the lowest LL via the fifth Haldane pseudopotential suppresses the composite Fermi liquid at $\nu=\frac{1}{2}$. It simultaneously favors the f-wave paired state and its two daughters. By contrast, the Pfaffian or anti-Pfaffian and their daughters are not favored. The sharp features in the top plot correspond to crossings between levels with different angular momentum.}
 \label{fig.v5}
\end{figure}

The wavefunction overlaps presented in Fig.~\ref{fig.v5} show that the ground state of the half-filled Landau level evolves from a composite Fermi liquid to the SU(2)$_2$ order at moderate $V_5$. Simultaneously, the two daughter states of this order also become favorable. By contrast, neither the Pfaffian order nor its daughters are stabilized by these interactions.

\bibliography{ref}

\end{document}